\documentclass[fleqn,10pt]{wlscirep}
\usepackage[utf8]{inputenc}
\usepackage[T1]{fontenc}
\title{The Heider balance and the looking-glass self}

\author[1]{Małgorzata J. Krawczyk}
\author[1]{Maciej Wołoszyn}
\author[1]{Piotr Gronek}
\author[1,*]{Krzysztof Kułakowski}
\author[2]{Janusz Mucha}
\affil[1]{AGH University of Science and Technology, Faculty of Physics and Applied Computer Science, al.~Mickiewicza~30, 30-059 Kraków, Poland.}
\affil[2]{AGH University of Science and Technology, Faculty of Humanities, ul.~Gramatyka~8a, 30-071 Kraków, Poland.}

\affil[*]{kulakowski@fis.agh.edu.pl}


\begin{abstract}
We consider the dynamics of interpersonal relations which leads to balanced states in a fully connected network. Here this approach is applied to directed networks with asymmetric relations, and it is generalized to include self-evaluation of actors, according to the 'looking-glass self' theory. A new index of self-acceptance is proposed: the relation of an actor to him/herself is positive, if the majority of his/her positive relations to others are reciprocated.  Sets of stable configurations of relations are obtained under the dynamics, where the self-evaluation of some actors is negative. Within each set all configurations have the same structure.
\end{abstract}
\begin{document}

\flushbottom
\maketitle
% * <john.hammersley@gmail.com> 2015-02-09T12:07:31.197Z:
%
%  Click the title above to edit the author information and abstract
%
\thispagestyle{empty}

%\noindent Please note: Abbreviations should be introduced at the first mention in the main text – no abbreviations lists. Suggested structure of main text (not enforced) is provided below.

\section*{Introduction}

%The Introduction section, of referenced text\cite{Figueredo:2009dg} expands on the background of the work (some overlap with the Abstract is acceptable). The introduction should not include subheadings.

Graphs are known as convenient mathematical representation of social networks for long time \cite{more,wasfa,scott}. In most applications, unweighted links have been used, with the only option that a link is present or not between each pair of nodes. Weak and strong links have been famously distinguished by Mark Granovetter \cite{gra1,gra2}; the difference was motivated by their roles in diffusion of information. Directed networks seem to be a natural next step in construction of sociological models, in particular when dealing with conflicts, social inequality and violence \cite{vala,york,staer,soy}. Yet asymmetric social relations have been modeled relatively rarely \cite{carl,lesk,cohen,bark,rami}.

Here we are interested in computational modeling of structural balance, which is a canonical example of an application of graphs in social science \cite{bonalu}. Briefly, the balance theory has been formulated by Fritz Heider in terms of positive and negative relations in triads \cite{fhei1,fhei}. Soon, a network of balanced triads has been shown to be equivalent to a balanced network \cite{hary}. On the other hand, the concept of balanced triads has found a psychological support in terms of cognitive dissonance \cite{lfest}. Our aim is to enrich the structural theory by a method of inference on self-evaluation \cite{cool} of actors, which they can reach on the basis of the inter-personal links between her/himself and other members of community.

In the next section we highlight sociological roots of the self-evaluation. In section 3 we clarify the role of the directions of links in the network triads. In two subsequent sections the scheme of calculation is described and the main results are given. The last but one section is devoted to discussion, where we demonstrate that our generic results are reproduced when sociometric data from literature are used. Short remarks on the applied numerical method close the text.

\section*{Self-evaluation by Cooley}

The issue of 'social relations' is, naturally,  one of foundations of sociology (including microsociology) and social psychology. Broadly understood 'interpretative social sciences' concentrate on proving that human attitudes (including emotions) and behaviors are reactions to the self-perceptions and self-images of the actors and their perceptions and images of their partners (human and non-human) rather than reactions to 'objective' human and non-human objects, processes and situations. Charles H. Cooley's contribution (see, e.g., \cite{cool}) is particularly significant in this context. His work is very influential until today in the broadly understood 'symbolic interactionism' (see, e.g., \cite{tzna,ghm,hblu,par}), in social psychology (see, e.g., \cite{zna,sch}), in 'sociology of individuals' (see, e.g., \cite{neli,jck}), even if practitioners of the latter do not quote Cooley directly. The influence of American classic's version of symbolic interactionism on today's 'sociology of emotions' (see, e.g., \cite{emo}) is highly visible. In this article, Cooley, as the founding father of original concepts and hypotheses still applied in various fields of social sciences, is a major source of inspiration.

Famous studies authored by Cooley were based on his analysis and interpretation of belles-lettres (particularly William Shakespeare's dramas) and his observations of social (including emotional) relations within his immediate family, mostly between his children. {\it If a boy [...] has any success, [...] he gloats over it [...]. He is eager to call in his friends [...], saying to them, 'See what I am doing! Is it not remarkable?' feeling elated when it is praised, and resentful or humiliated when fault is  found with it} (\cite{cool}, p. 178).

Let us quote a significant piece of Cooley's book 'Human Nature and the Social Order': {\it A self-idea [...] seems to have three principal elements: the imagination of our appearance to the other person; the imagination of his judgment of that appearance, and some sort of self-feeling, such as pride or mortification. [...] The thing that moves us to pride or shame is not the mere mechanical reflection of ourselves, but an imputed sentiment, the imagined effect of this reflection upon another's mind. This is evident from the fact that the character and weight of that other, in whose mind we see ourselves, makes all the difference with our feeling. We are ashamed to seem evasive in the presence of a straightforward man, cowardly in the presence of a brave one, gross in the eyes of a refined one, and so on. We always imagine, and in imagining share, the judgments of the other mind. A man will boast to one person of an action – say some sharp transaction in trade – which he would be ashamed to own to another} (\cite{cool},pp. 184-5). 

From a sociological perspective, our results indicate that negative self-evaluations come from unreciprocated relations. We dare say that this statement links the concept of Heider balance with the idea of 'looking-glass self' of C. H. Cooley (\cite{cool}, p.152). Sometimes reported in the form of an aphorism: {\it each to each a looking glass/ reflects the other that doth pass} \cite{cool}, the idea highlights the role of society in formation of human identity. Namely, opinion of an individual on herself/himself is formed by perceived opinions of other people.

\section*{Equations of motion}

Consider a fully connected network of $N$ nodes and $K=N(N-1)$ links between them, $x_{ij}=\pm1$. If links are set to be symmetric ($x_{ij}=x_{ji}$), the condition of structural balance is that for each triad $(i,j,k)$ of nodes the links between them obey the rule $x_{ij}x_{jk}x_{ki}=+1$. It is known \cite{hary}, that in this case the whole network is split in two parts, where $x_{ij}=+1$ if nodes $i,j$ belong to the same part, otherwise $x_{ij}=-1$. This split is termed as 'structural balance' or 'Heider balance' (HB).  Putting this in terms of friendly and hostile relations, all relations within each group are friendly, and all relations between groups are hostile. This means in particular, that the hostile relation is not transitive: an enemy of my enemy is my friend.

If the link variables 
$x_{ij}$ are allowed to be real, the following set of differential equations of time evolution leads generically to HB \cite{my05,mar}:
\begin{equation}
\frac{dx_{ij}}{dt}=(1-x_{ij}^2)\sum_{k}^{N-2}x_{ik}x_{kj}
\label{e1}
\end{equation}
where $k\ne i, k\ne j$ in the sum. The rationale of Eqn. (\ref{e1}) is as follows. Once $k$ is either a common friend of $i$ and $j$ or their common enemy, the product $x_{ik}x_{kj}$ is positive and contributes to an increase of $x_{ij}$ in time. On the other hand, if $k$ is a friend of $i$ but an enemy of $j$ (or the opposite), the product $x_{ik}x_{kj}$ is negative and contributes to a decrease of $x_{ij}$. The factor $1-x_{ij}^2$ keeps the relation $x_{ij}$ in the finite range $[-1,+1]$; this is the price we pay for the lack of analytical solution of Eqn. (\ref{e1}) \cite{mar}. We note that discrete algorithms leading to HB have been formulated also for the symmetric case \cite{r1,r2}.

When the condition of symmetry is released, the order of indices does matter. Basically, the inference of actor $i$ about her/his relation to $j$ could be modeled in several ways of ordering; several possible types of directed and signed triads have been considered \cite{carl,lesk,ramb,mol,sad}. Our choice to keep the ordering as in Eqn. (\ref{e1}) \cite{my0,my2} is driven by the lack of transitivity, mentioned above. In terms of social relations, the question on transitivity \cite{wasfa} is: once actor A wonders about his feeling about B, how important is that A dislikes C and C dislikes B? We can imagine that the issue is less important when B is a product \cite{lesk}, on the contrary to the case when B is an enemy of a friend. We note also that another ordering $x_{ik}x_{jk}$, symmetric vs an exchange $i$ and $j$, would drive the system to symmetric relations, and therefore is out of interest here. Finally, we should add that we treat the character of all relations as known to all actors, what is appropriate only when dealing with small groups; more general picture has been discussed by Carley and Krackhardt \cite{carl}.

 The index $F_i$ of self-evaluation of each actor ($i=1,2,...,N$) is calculated as  

 \begin{equation}
 F_i=\frac{1}{2}\sum_k (1+x_{ik})x_{ki} .
 \label{e2}
 \end{equation}
 This form of $F_i$ is appropriate as long as we are interested only in the opinions of actors $k$ liked by $i$ ($x_{ik}=+1$); those whom $i$ dislikes ($x_{ik}=-1$) do not contribute to $F_i$. Here we are motivated by the concept od significant others \cite{sio}. Basically, $-N < F_i <N$.

 \section*{Calculations}

Some initial configurations lead to balanced states which are necessarily symmetric. In a symmetric state $F_i \geq 0$ for each actor $i$; $F_i$ is just the 
 number of friends of $i$, and this friendship is reciprocated. Some others lead to asymmetric states, which cannot be balanced; here we are interested in those states where $F_i<0$ for some actors $i$. Among those liked by her/him, such an actor has more enemies than friends; $F_i$ is just the difference between friends and enemies of this kind. 
 
 At this stage, the outcome of the calculation is a set of $K(N)$ matrices $x_{ij}$. We prefer to keep $N$ odd to evade the cases where the right side of Eq. ({\ref{e1}}) is equal to zero in a stationary state. We have got results for $K=10^4$ for each $N=7,9,11,41,55$ and $77$; for $N=99$, the dynamic system is found to be less stable numerically. For each $N$, the matrices are classified as equivalent to unlabeled graphs of classes of nodes. In short, the algorithm is as follows \cite{mjk1,mjk2}:
 
 \begin{enumerate}
 \item for each node $i$, the number $M(i)$ of nodes is found such that $x_{im}=+1$, and the number $L(i)$ of nodes is found such that $x_{li}=+1$ ;
 \item nodes of the same $M,L$ are provisionally classified as belonging to the same class;
 \item neighbours of nodes in the same class are checked if they belong to the same classes;
 \item if not, more fine classes are introduced;
 \item steps 3 and 4 are repeated until the condition 3 is true.
 \end{enumerate}
 
 The resulting graph is encoded as the matrix of relations between the classes plus information on the numbers of nodes in each class. We note that all nodes in the same class have the same numbers of neighbors in the same classes; the same applies to neighbors of neighbors etc \cite{mjk1,mjk2}. Obviously, all nodes $i$ in the same class have the same value of $F_i$.

\section*{Results}

%Up to three levels of \textbf{subheading} are permitted. Subheadings should not be numbered.

\begin{figure}[!htb]
\begin{center}
\includegraphics[width=\linewidth]{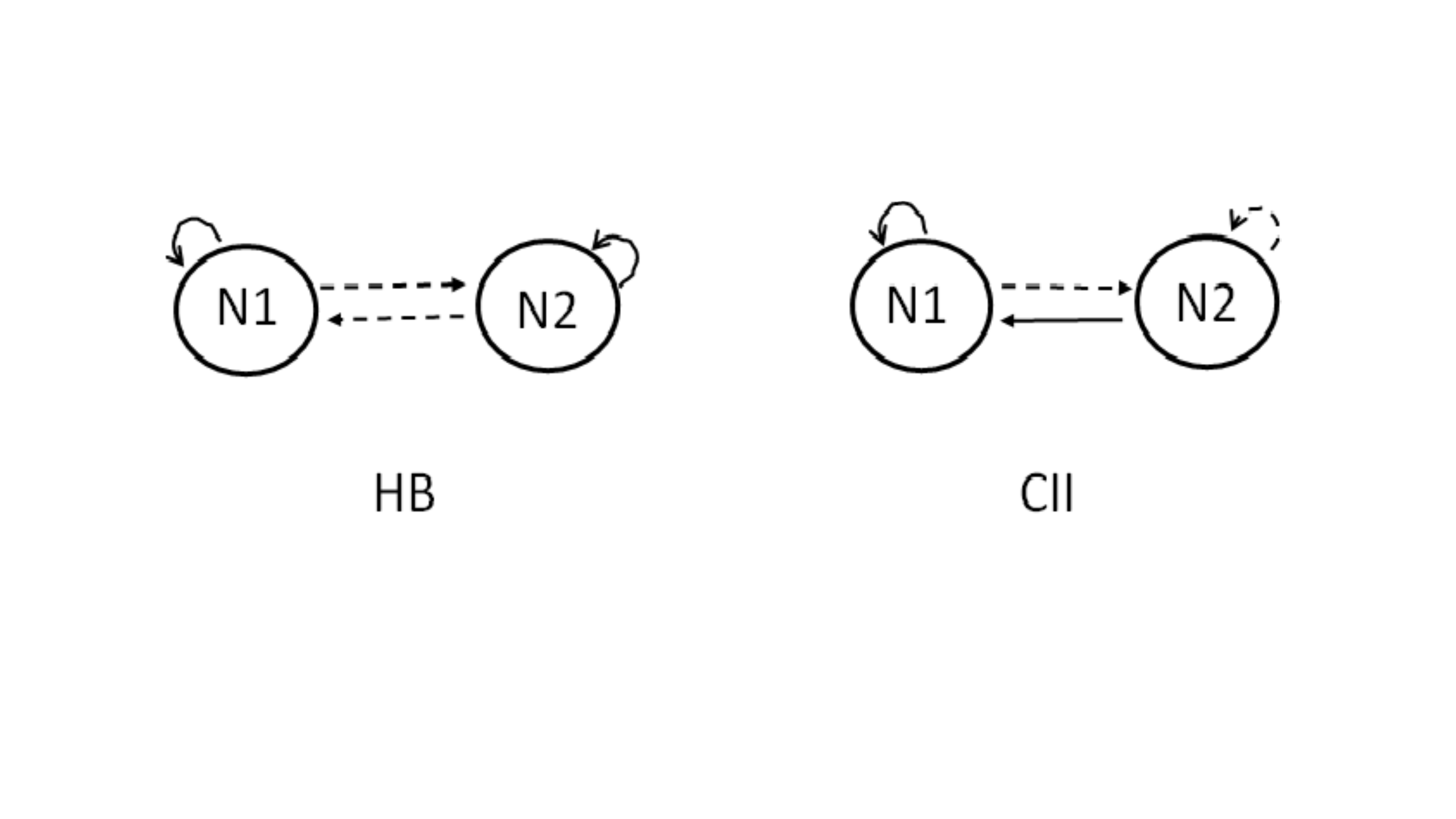}
\caption{The generic graphs of two classes. The graph noted as HB shows the usual form of the Heider balance: the classes contain $N1$ and $N2$ nodes (actors), with all relations within the same class friendly, all relations between the classes hostile, and all links of this graph symmetric. The graph noted as CII contains class 1 with friendly internal relations and class 2 with internal relations hostile. The relations of those from $N1$ towards those from $N2$ are hostile (dashed arrow), and the relations of those from $N2$ towards those from $N1$ are friendly (continuous arrow).}
\label{f0}
\end{center}
\end{figure}

The results on the matrices of relations between classes can be written in the form of graph of classes. To make it more clear, we provide two examples in Fig. \ref{f0}. There, a continuous (dashed) arrow from class A to class B means a positive (negative) relation of members of A towards class B. The same applies to arrows from A to A; in this case the relation is of the group members towards other members of the same group; the relation $F_i$ of an actor to him/herself is not marked in the figures. On the left there (HB), a classical case of the Heider balance is shown; all links are symmetric, then there are no nodes with $F_i$ negative, and the graph is shown for completeness only. On the right side of the figure, the graph CII contains two classes: class 1 with internal relations friendly, and class 2 with internal relations hostile. The stability conditions for a friendly relation $x_{ij}$ is that the sum $\sum x_{ik}x_{kj}$ (as in Eq. (\ref{e1})) should be positive. For the internal links within the class 1 this sum is equal $N1-2-N2$, hence the stability in this case yields $N1>N2+2$. For the internal links within the class 2, the appropriate sum should be negative, then their stability is assured if $N2-2-N1<0$ - a less demanding condition. Accordingly, the negative link from class 1 to class 2 is stable if $-(N1-1)+N2<0$, and the positive link from class 2 to class 1 is stable if $-(N2-1)+(N1-1)>0$. Summarizing this thread, the stability condition for the graph CII is that $N1>N2+2$. The self-evaluation index $F_1=N1-1$ for the class 1, and $F_1=-N1$ for the class 2. Yet we note that the graphs CII are more rare, than the others.

\begin{figure}[!htb]
\begin{center}
\includegraphics[width=\linewidth]{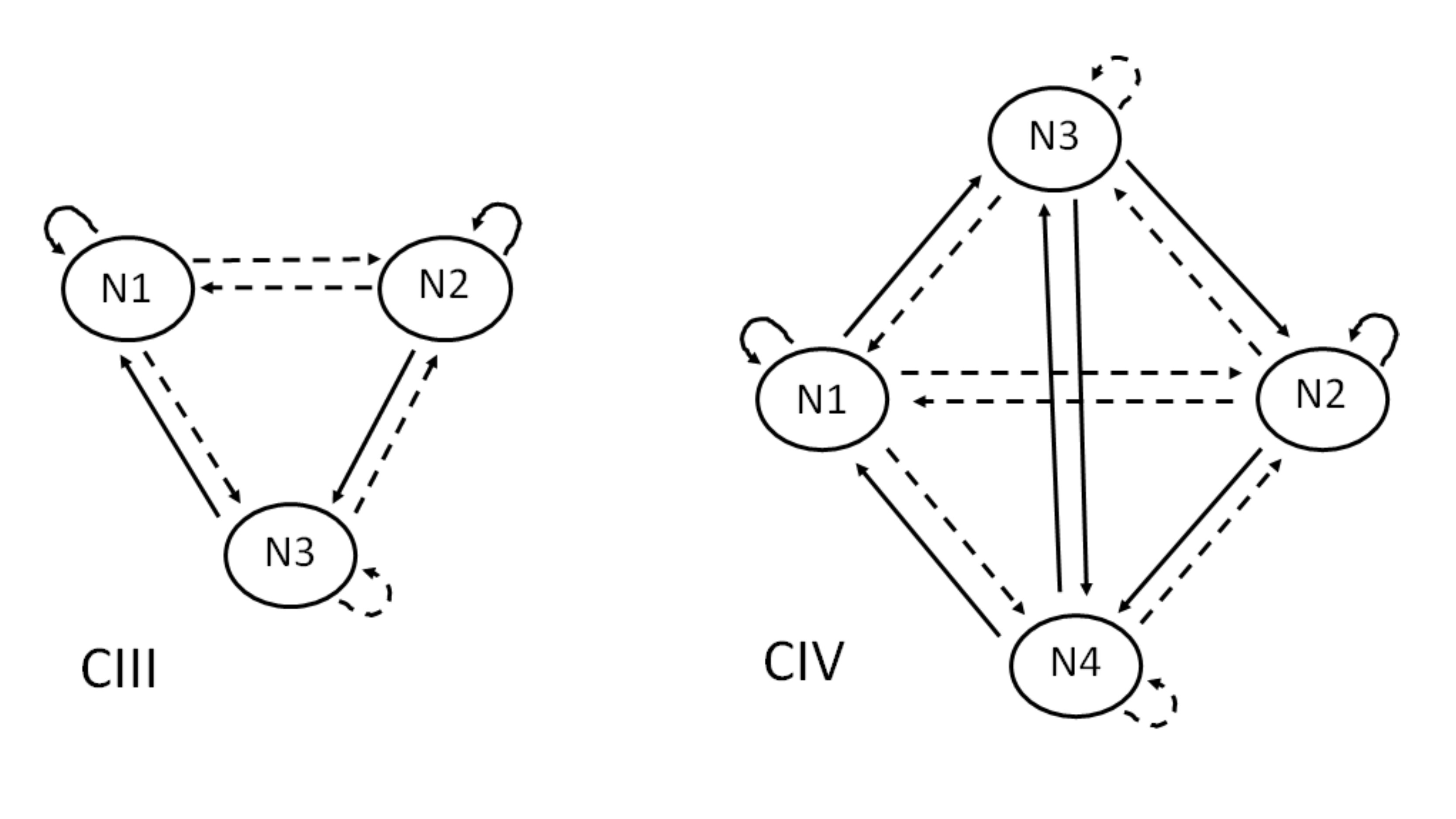}
\caption{The generic graphs of three (CIII) and four (CIV) classes. N1-N4 mean the numbers of nodes (actors) in a given class. Continuous arrows from N1 to N2 mean friendly relations of $N1$ towards $N2$ (members of 1 like those in 2), and dashed arrows mean hostile relations. The same applies to self-directed arrows: a dashed self-directed arrow means that all members of a given class dislike each other and themselves.}
\label{f1}
\end{center}
\end{figure}

Numerical results indicate that two kinds of configurations of graphs of classes appear much more frequently than others. These two kinds can be classified as CIII and CIV, as they are composed of three and four classes, respectively. Both CIII and CIV appear for different numbers $N$ of actors in particular classes. These graphs are shown in Fig. {\ref{f1}}, with the same meaning of continuous and dashed arrows.

For the graph CIII, the stability condition $x_{ij}=sign(\sum_k x_{ik}x_{kj})$ is equivalent to the following: $N1+N2>N3+2$ for links within class 1, within 2, from 1 to 2 and back, and within class 3. For other links (from 1 to 3 and back, and from 2 to 3 and back), the condition is weaker: $N1+N2>N3$. Similarly, for CIV the most demanding condition is $N1+N2>N3+N4-2$. As we see in Table \ref{t1}, all the obtained links are stable. In the last column of Table \ref{t1} we show the obtained statistics of CIII and CIV for $N=7$. The results indicate that asymmetric solutions appear in about 73 percent of our sample of $10^4$ networks.

\begin{table}
\centering
\caption{Most frequent configurations of graphs of classes for CIII ($N4=0$) and CIV ($N4>0$) for $N=7$. 
$N1\div N4$ mean the numbers of nodes (actors) in a given class 1$\div$4, and $\#$ counts how many times the graph appeared at the end of evolution in the sample of $10^4$ networks. The structure of graphs is shown in Fig. \ref{f1}. For CIII, configurations are shown for $\#>100$, and for CIV - for $\#>5$.}
\begin{tabular}{|c|c|c|c|c|}
\hline
N1&N2&N3&N4&$\#$\\
\hline
3&3&1&0&1465\\
\hline
4&2&1&0&1125\\
\hline
2&4&1&0&1111\\
\hline
3&2&2&0&474\\
\hline
2&3&2&0&462\\
\hline
1&5&1&0&443\\
\hline
5&1&1&0&422\\
\hline
1&4&2&0&226\\
\hline
4&1&2&0&224\\
\hline
2&3&1&1&929\\
\hline
4&1&1&1&442\\
\hline
\end{tabular}
\label{t1}
\end{table}

\begin{figure}[!hptb]
\begin{center}
\includegraphics[width=\linewidth]{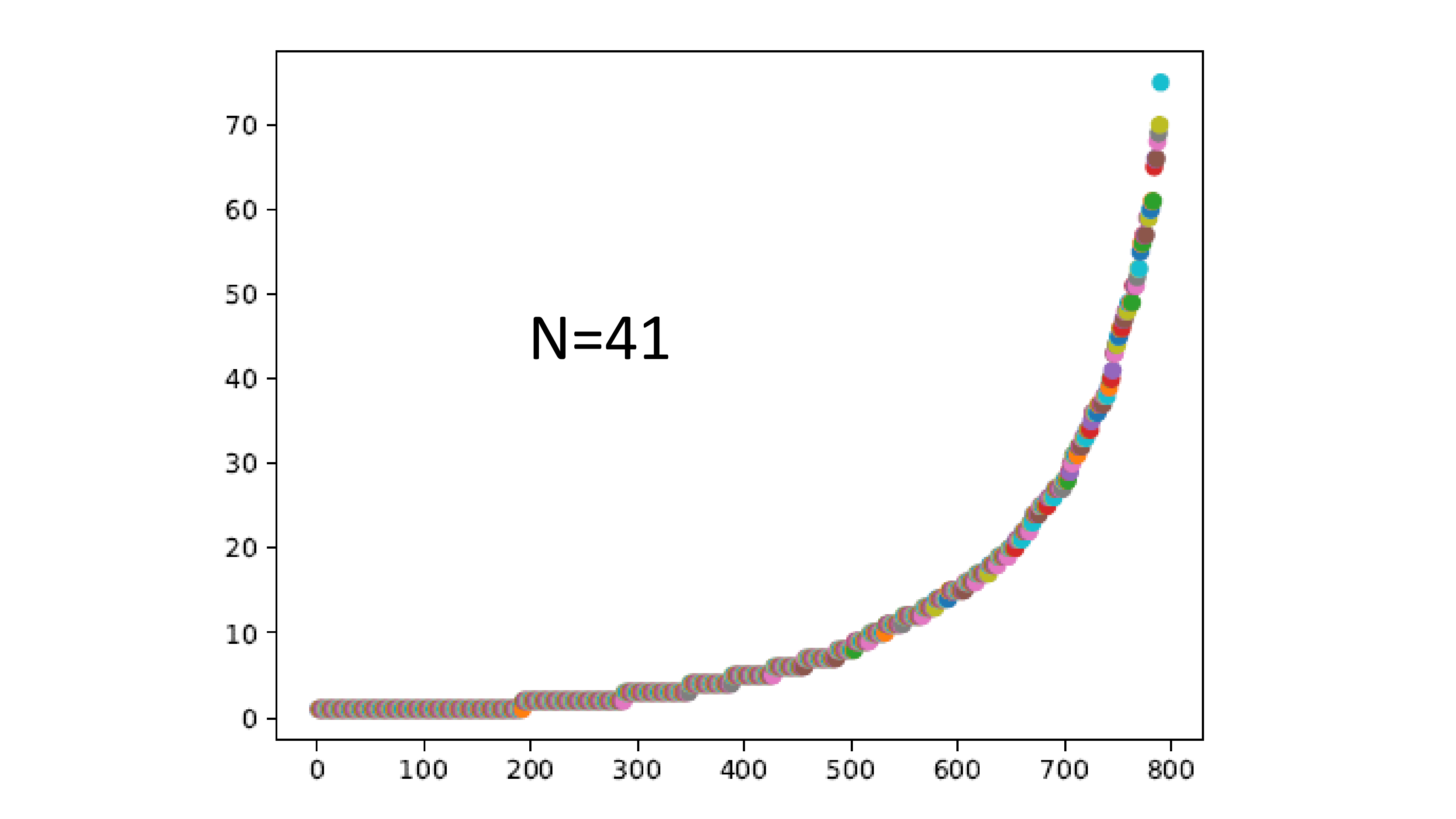}
\caption{Frequency $\#$ of different graphs CIV for $N=41$. By different graph we mean that the numbers $N1\div N4$ are different. The graphs are ordered with increasing frequency; higher rank is assigned to those which appear more frequently.}
\label{f2}
\end{center}
\end{figure}

\begin{figure}[!hptb]
\begin{center}
\includegraphics[width=\linewidth]{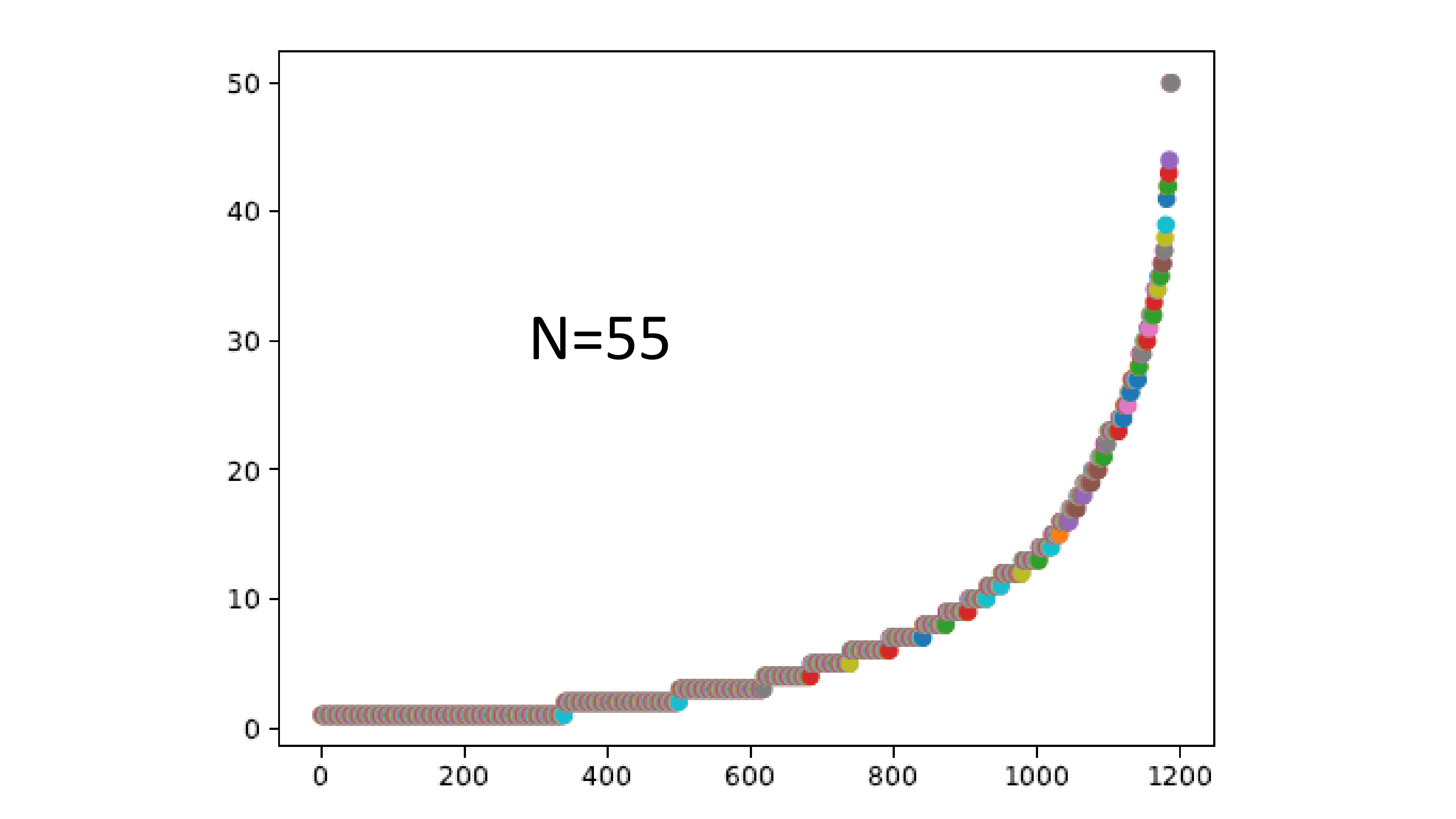}
\caption{Frequency of different graphs CIV for $N=55$. By different graph we mean that the numbers $N1\div N4$ are different. The graphs are ordered with increasing frequency; higher rank is assigned to those which appear more frequently.}
\label{f3}
\end{center}
\end{figure}

\begin{figure}[!hptb]
\begin{center}
\includegraphics[width=\linewidth]{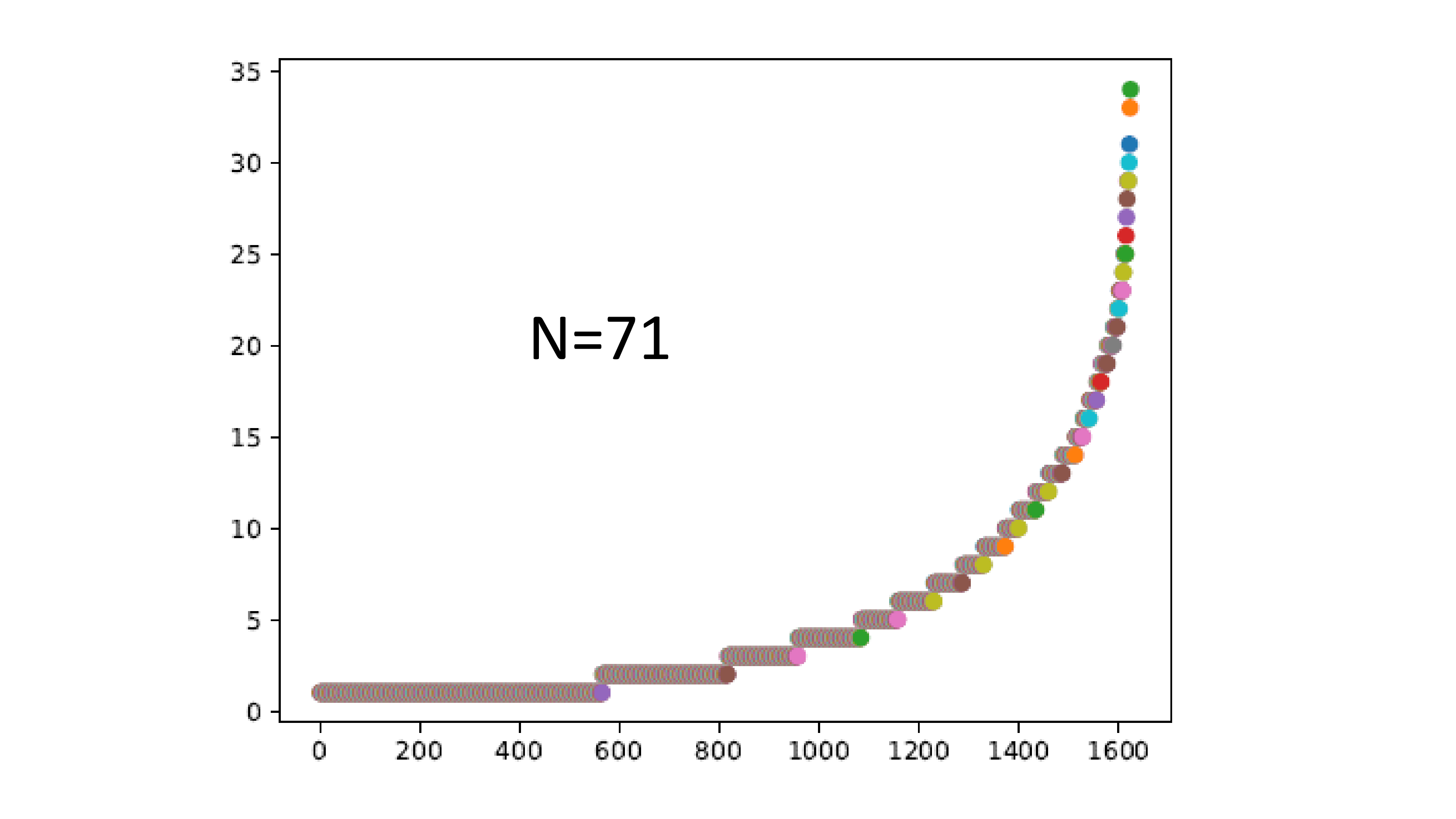}
\caption{Frequency of different graphs CIV for $N=71$. By different graph we mean that the numbers $N1\div N4$ are different. The graphs are ordered with increasing frequency; higher rank is assigned to those which appear more frequently.}
\label{f4}
\end{center}
\end{figure}

For larger values of $N$, the number of configurations is larger, but again CIII and CIV appear more frequently. For $N=9$, the number of actors in the graphs of classes of leading frequencies are listed in Table \ref{t2}. The sum of the right column covers more than 78 percent of the sample. For $N=11$, we got 3686 graphs CIII and 5120 graphs CIV, what makes 88 percent of the sample. For $N=41$ the population changes: 6974 graphs CIV (only those with frequency $n$ not less than 10) and only 10 single graphs CIII. For $N=55$ and $N=71$ there are no graphs CIII at all. This effect is presumably a consequence of the fact that the number of partitions of $N$ into four non-zero sets increases with $N$ much quicker, than the number of partitions of $N$ into three non-zero sets. Basically, these numbers are known as the Stirling numbers of second kind, $S(N,4)$ and $S(N,3)$. The ratio $S(N,4)/S(N,3)$ is 1.2 for $N=7$, 2.6 for $N=9$, 5.1 for $N=11$ but about $3\times 10^4$ for $N=41$ \cite{mohr}. This calculation is not exact because the stability conditions are ignored there, but even a very rough evaluation explains the observed lack of CIII for $N>40$. In Figs. \ref{f2}, \ref{f3} and \ref{f4} the frequencies of various graphs CIV (with different numbers $N1\div N4$) are shown. We can conclude that the percentage of asymmetric graphs is meaningful.

\begin{table}
\centering
\caption{Most frequent configurations of graphs of classes for CIII ($N4=0$) and CIV ($N4>0$) for $N=9$. The notation is as in Table \ref{t1}. For CIII, configurations are shown for $\#>150$, and for CIV - for $\#>3$.}
\begin{tabular}{|c|c|c|c|c|}
\hline
N1&N2&N3&N4&$\#$\\
\hline
4&4&1&0&721\\
\hline
5&3&1&0&610\\
\hline
3&5&1&0&597\\
\hline
4&3&2&0&547\\
\hline
3&4&2&0&545\\
\hline
5&2&2&0&347\\
\hline
2&5&2&0&327\\
\hline
2&6&1&0&294\\
\hline
6&2&1&0&276\\
\hline
3&3&3&0&172\\
\hline
3&4&1&1&1084\\
\hline
5&2&1&1&664\\
\hline
3&3&1&2&435\\
\hline
4&2&1&2&375\\
\hline
4&2&2&1&353\\
\hline
1&6&1&1&196\\
\hline
5&1&1&2&139\\
\hline
5&1&2&1&138\\
\hline
\end{tabular}
\label{t2}
\end{table}

Another question is the sign of $F_i$ in the obtained classes. Here we are interested in the cases where $F_i<0$. For CIII, this condition is equivalent to: $N1<1$, $N2<N3+1$ and $N1>0$ for the actors in the class 1, 2 and 3, respectively. As we can infer from the data in Table \ref{t1} for $N=7$,  $F_i$ is always negative only for actors in class 3. The self-feeling $F_i$ of the remaining nodes is sometimes zero, sometimes positive, sometimes it can be negative as well. Similarly, for CIV the values of $F_i$ are equal to $N1-N3-1$, $N2-N4-1$, $N4-N2$ and $N3-N1$ for actors in nodes 1,2,3 and 4, respectively. As we can infer from the stability condition given above ($N1+N2>N3+N4+2$), $F_i$ is negative for most actors in the classes $N3$ and $N4$.

% \subsection*{Subsection}
% 
% Example text under a subsection. Bulleted lists may be used where appropriate, e.g.
% 
% \begin{itemize}
% \item First item
% \item Second item
% \end{itemize}
% 
% \subsubsection*{Third-level section}
%  
% Topical subheadings are allowed.

\section*{Discussion}

%The Discussion should be succinct and must not contain subheadings.

Our numerical results indicate that:
\begin{itemize}
\item the evolution equation (\ref{e1}) frequently drives the system to asymmetric states with structure CIII or CIV as shown in Fig. \ref{f1}, 
\item in small systems ($N \approx 10$) graphs of classes CIII and CIV are abundant, while in larger systems ($N > 40$) graphs CIII disappear,
\item the stability conditions demand that the classes 1 and 2 (with positive internal relations) contain more actors than the classes 3 and 4,
\item for actors in the classes 3 and 4, negative values of $F_i$ are quite typical.
\end{itemize}

As it was demonstrated in Refs. \cite{my05,mar}, in the case of symmetric relations ($x_{ij}=x_{ji}$) the generic stable solution of the Eq.(\ref{e1})
is balanced in the Heider sense. An example of such solution is shown in Fig. \ref{f1} (HB) in the form of a graph of classes. Recall that there are two classes, one with $N1$ actors and one with $N2$. Here we show that when the condition of symmetry is released, the graphs of classes CIII and CIV
(shown in Fig. \ref{f1}) are most probable solutions of Eq. (\ref{f1}). When the number of actors $N$ increases, the frequency of CIV prevails. While the numbers of actors $N1, N2, N3, N4$ vary, the structure of the graph of classes remains unchanged. We dare say that the structure of CIII and foremost CIV with its specific symmetry are counterparts of the balanced state for asymmetric relations. What we do not know is, if there is no more solutions of this kind. Recall that for symmetric relations and $N=9$, a stable state of three triads has been found, which is different from a classical balanced partition into two groups \cite{r1}. At this stage we cannot state that there is no other asymmetric and stable states, different from CIII and CIV.  

To comment the structure of the graphs, the graph CII could represent peer rejection \cite{perr} of those in class 2 by those in class 1 and by other members of the class 2. The class 3 in CIII can be assigned to a person (or persons) at intermediate position between two mutually hostile groups. She/he aspires to belong to one of them, but is not accepted there; the aspiration makes her/him unwilling to join the other group, where she/he would be accepted. This scheme reminds some fictitious characters known in belles-lettres, as Werther \cite{goe} or Julian Sorel \cite{sten}. In general, the structure of CIII should be visible when characterizing social positions of mobile individuals, aspiring to move between mutually hostile groups. A similar scheme of relations has been identified in ethnic conflicts, where higher educated immigrants perceive more discrimination that their less educated compatriots. The phenomenon is known as 'paradox of integration' (see \cite{teij,vroo,dix,tols} and references therein). Going to the graph IV, we can imagine two mutually hostile groups 1 and 2 (as usually in the balanced state) plus one or two relatively small groups 3 and 4. This structure is found to be stabilized by the same process of removal of cognitive dissonance, which - if backed by the condition of symmetric relations - leads to the Heiderian balanced state. Here we apply the Cooley theory of looking-glass self to infer on the self-evaluation of actors in the minor groups 3 and 4. Our main result here is to indicate that the self-feeling is clearly worsened there by the structure of relations.

As it can be inferred from our numerical results, the graphs CIII and CIV appear to be generic. Therefore we can expect that they should be present in some sociometric data. This expectation has been verified by using the Sampson data \cite{uci}. Namely, four pairs of non-symmetric, valued matrices (X,Y)=(SAMPLK,SAMPDLK), (SAMPES,SAMPDES), (SAMPIN,SAMPNIN), (SAMPPR,SAMPNPR) \cite{uci} were used to form four signed and valued non-symmetric matrices (X-Y)/5. The difference measures the positive minus the negative relations, and the factor $1/5$ was used to keep the relations in the range 
$(-1,1)$. These four matrices have been used as the initial values of $x_{ij}$. We have found that only one pair (SAMPLK-SAMPDLK)/5 was dense enough to produce a fully connected graph by means of Eqn.(\ref{e1}). The structure of this graph has been found to be identical with CIII. There, the class 1 contains ten actors (1,2,4,5,6,7,11,12,14,16), the class 2: four actors (13,16,17,18), and the class 3 also four (3,8,9,10), with enumeration as the rows and columns of the matrices given in \cite{uci}. Two other matrices (SAMPES-SAMPDES)/5 and (SAMPIN-SAMPNIN)/5 appeared to be dense enough, i.e. their evolution lead to fully connected graphs, only if the first row and column was removed; the first actor was apparently less connected than the others. With this cut, the pair (SAMPIN-SAMPNIN)/5 has been found to lead to CIII, with the composition $N1=7,N2=9,N3=1$. The matrix (SAMPES-SAMPDES)/5 gave CIV with the composition $N1=5,N2=7,N3=3,N4=2$. The matrix (SAMPES-SAMPDES)/5 gave CIV with the composition $N1=5,N2=7,N3=3,N4=2$, and the same composition has been produced by (SAMPLK-SAMPDLK)/5 as the initial state. The fact that both CIII and CIV are reproduced from the well-known set \cite{uci} suggests that these graphs are generic also in other sociometric data.

To conclude, the former theory of removal of cognitive dissonance \cite{my05} has been extended here to include asymmetric relations. Our results indicate, that the time 
evolution starting from generic initial conditions leads in most cases to the graphs of classes CIII and CIV. According to the Cooley theory, the structure of these graphs is particularly harmful for actors at some positions in these graphs, indicated above. On the other hand, this structure can be identified in field experiments by means of standard sociometric methods. Our results can be of interest for scientists working on conflicts in groups, and for teachers in classes where conflicts appear.

\section*{Methods}

%Topical subheadings are allowed. Authors must ensure that their Methods section includes adequate experimental and characterization data necessary for others in the field to reproduce their work.

For each sample out of a set of $K$ fully connected networks of size $N$, initial conditions for $N(N-1)$ links $x_{ij}(t=0)$ have been chosen randomly from a homogeneous distribution $\rho(x)=1/2$ for $-1<x<1$, zero otherwise. Next, the time evolution of $x_{ij}$ has been initialized according to Eqn.~(\ref{e1}).
The resulting set of differential equations has been solved numerically using the \emph{Mathematica} software for a sample of $K=10^4$ networks for each $N$.
Typically, the system has ended at one of hypercube corners where $x_{ij}=\pm 1$ (with the relative accuracy $10^{-8}$ serving as the stop condition of the simulation) for each pair $i,j$.
The convergence to $x_{ij}=\pm 1$ has been reached in 97\% of the cases for $N=7,9,11$, in 96\% cases for $N=41$, in 95\% cases for $N=55$,  and in 93\% cases for $N=77$.
For $N=99$, the numerical solution of the system of differential equations has proved to be unstable, with at least 55\% of the cases not converging.

%%%%%%%%%%%%%%%%%%%%%%%%%%%%%%%%%%%%%%%%%%%%%%%%%%%%%%%%%%%%%%%%%%%%%%%%%%%%%%

% \bibliography{sample}
%
% \noindent LaTeX formats citations and references automatically using the bibliography records in your .bib file, which you can edit via the project menu. Use the cite command for an inline citation, e.g.  \cite{Hao:gidmaps:2014}.
% 
% For data citations of datasets uploaded to e.g. \emph{figshare}, please use the \verb|howpublished| option in the bib entry to specify the platform and the link, as in the \verb|Hao:gidmaps:2014| example in the sample bibliography file.

%%%%%%%%%%%%%%%%%%%%%%%%%%%%%%%%%%%%%%%%%%%%%%%%%%%%%%%%%%%%%%%%%%%%%%%%%%%%%%

\section*{Acknowledgements}

%Acknowledgements should be brief, and should not include thanks to anonymous referees and editors, or effusive comments. Grant or contribution numbers may be acknowledged.

This work was partly supported by the Faculty of Physics and Applied Computer Science (11.11.220.01/2) and by the Faculty of Humanities (11.11.430.158) AGH UST statutory tasks within subsidy of Ministry of Science and Higher Education. Support by the facilities of the PL-Grid infrastructure is also acknowledged.

\section*{Author contributions statement}

K.K. designed the research, M.J.K, M.W. and P.G. performed the simulations, K.K. analysed the results, K.K., J.M. and M.W. wrote the text.  All authors reviewed the manuscript. 

\section*{Additional information}

%To include, in this order: \textbf{Accession codes} (where applicable); \textbf{Competing interests} (mandatory statement). 

%The corresponding author is responsible for submitting a \href{http://www.nature.com/srep/policies/index.html#competing}{competing interests statement} on behalf of all authors of the paper. This statement must be included in the submitted article file.

\textbf{Competing interests}:
The authors declare no competing interests.

\end{document}